\documentclass[letterpaper,onecolumn,12pt]{article}

\usepackage[T1]{fontenc}  
\usepackage{multicol}
\usepackage{amsmath,amssymb}
\usepackage{gensymb}
\usepackage{times}
\usepackage{fullpage}
\usepackage[figure,vlined,linesnumbered]{algorithm2e}
\usepackage{color}
\usepackage{rotating}
\usepackage{soul}
\usepackage{refcount}
\usepackage{helvet}
\usepackage{url}
\usepackage{refcount}
\usepackage[numbers,sort]{natbib}
\usepackage{inconsolata}
\usepackage{subfigure}
\usepackage{float}
\usepackage{enumitem}
\usepackage{multirow}
\usepackage{fancyhdr}
\usepackage{verbatim}
\usepackage{array}
\usepackage{pdfsync}
\usepackage[breaklinks,colorlinks,linkcolor=black,citecolor=black,urlcolor=black]{hyperref}
\usepackage{cleveref}
\usepackage[export]{adjustbox}
\usepackage{titling}
\usepackage{titlesec}
\usepackage{lipsum}

\newcommand{\ane}{E}
\newcommand{\aneb}{\tilde{\ane}}
\renewcommand{\mod}{\text{ mod }}

{\hspace*{\fill}$\Box$\par\vspace{4mm}}

\titleformat{name=\section}{\normalfont\Large\bfseries}{}{0pt}{}
\titleformat{name=\subsection}{\normalfont\large\bfseries}{}{0pt}{}
\titleformat{name=\subsubsection}{\normalfont\normalsize\bfseries}{}{0pt}{}

\title{An authenticated and secure accounting system for international emissions trading}

\begin{document}

\date{}

\newcommand{\oneway}{f}
\newcommand{\groupadd}{\phi}
\newcommand{\groupmulf}{\psi}
\newcommand{\groupmul}[1]{\groupmulf^{#1}}
\newcommand{\infp}{{\cal O}}

\begin{titlepage}
{\noindent\LARGE\bf\thetitle}

\newcommand\blfootnote[1]{%
  \begingroup
  \renewcommand\thefootnote{}\footnote{#1}%
  \addtocounter{footnote}{-1}%
  \endgroup
}

\bigskip

\begin{flushleft}\large
Chenxing Li\textsuperscript{1,$\dagger$},
Yang Yu\textsuperscript{1,3,$\dagger$},
Andrew Chi-Chih Yao\textsuperscript{1,3,*},
Da Zhang\textsuperscript{2,*},
Xiliang Zhang\textsuperscript{2,*}
\end{flushleft}

\bigskip

\noindent
\begin{enumerate}[label=\textbf{\arabic*}]
\item Institute for Interdisciplinary Information Since,Tsinghua University
\item Institute of Energy, Environment and Economy, Tsinghua University
\item Shanghai Qi Zhi Institute
\end{enumerate}

\bigskip

\blfootnote{$^\dagger$ These authors contribute equally.}
\blfootnote{$^*$  Correspondenance authors. E-mail: andrewcyao@tsinghua.edu.cn (Andrew Yao), zhangda@tsinghua.edu.cn\\ (Da Zhang), zhang\_xl@tsinghua.edu.cn (Xiliang Zhang)}

\bigskip

\section*{Abstract}
Expanding multi-country emissions trading system is considered as crucial to fill the existing mitigation gap for the 2\degree C climate target. Trustworthy emissions accounting is the cornerstone of such a system encompassing different jurisdictions. However, traditional emissions measuring, reporting, and verification practices that support data authenticity might not be applicable as detailed data from large utilities and production facilities to be covered in the multi-country emissions trading system are usually highly sensitive and of severe national security concern. In this study, we propose a cryptographic framework for an authenticated and secure emissions accounting system that can resolve this data dilemma. We demonstrate that integrating a sequence of cryptographic protocols can preserve data authenticity and security for a stylized multi-country emissions trading system. We call for more research to promote applications of modern cryptography in future international climate governance to build trust and strengthen collaboration.

\end{titlepage}

\thispagestyle{empty}
The multi-country emissions trading scheme (ETS) is a crucial mechanism for promoting international cooperation on climate change mitigation and enhancing countries’ ambition. The multi-country ETS can reduce mitigation costs by allowing countries to direct investment in the cheapest emissions reduction opportunities. The EU-ETS has demonstrated an initial success of the scheme in coordinating multilateral efforts to reduce CO$_2$ emissions \cite{Bayer2020}.

A fundamental trust problem has hindered the further expansion of the multi-country ETS to more jurisdictions. In a typical multi-country ETS setting, covered firms in different countries trade their emissions permits to meet the requirement that the same number of permits as their emissions should be submitted by the end of each compliance cycle (e.g., one year). Therefore, trustworthy reporting of firm-level emissions data is a prerequisite of a well-functioning ETS. However, once emissions are traded with a price, firms have incentives to misreport or even manipulate their emissions data \cite{Duflo2013}, jeopardizing the trust in the ETS. In the context with a central authority presented (e.g., a national government or supranational environmental regulator), existing studies have shown that implementing (at least partially) random emissions verification on a subset of firms is an effective solution \cite{Duflo2013,Duflo2018,Zhang2019}. Authorized trustworthy verifiers can conduct on-site visits to validate firms’ self-reported data by examining granular emissions (and production) data and checking the operating status of metering equipment that generate the granular data. Hence, an effective measuring, reporting, and verification (MRV) system that exposes firms to random verification could serve as a valid threat to prevent them from misreporting data.

However, random verification entails major challenges in the context of a typical multi-country ETS. Although countries may agree to allow a subset of firms to receive verification, a data dilemma emerges as there is a trade-off between data authenticity and data security. On the one hand, granular emissions data that meet certain international data quality standards are required to stay untampered after being generated and stand ready for possible verification. On the other hand, these granular data are highly sensitive and usually contain confidential information related to commercial secrets and national security, because large emissions facilities to be covered by the multi-country ETS are usually strategic units related to energy infrastructure or advanced technology development. Hence, it will be considered unacceptable by most firms and countries if the granular data prepared for the possible verification---in many cases firm-level aggregated emissions data as well---face the risk of unauthorized exposure.

Ensuring true randomness of the verification itself, however, is also a challenge. Some firms (out of security or reporting quality concerns) always prefer not to be selected and have the incentives to influence the randomization process. Without having complete confidence in the firm selection process that can guarantee true randomness against potential interventions by firms, countries, or other adversary parties, firms and countries involved in the verification cannot fully cooperate without any disputes. 

To address these challenges, we introduce modern cryptography in the field of global climate governance. Modern cryptography has developed mathematical theories~\cite{RSA,DH,yao1982theory,goldwasser1984probabilistic} and algorithms~\cite{katz2014introduction,goldreich2009foundations} to address communication challenges that involve information authenticity and security, building trust and ensuring security inherently in many settings. Beyond hiding information from misuse, cryptography can also achieve information binding that prevents parties from altering their submitted data and ensure information commitment~\cite{blum1983coin,damgaard1998commitment}. For instance, the famous cryptocurrency, Bitcoin~\cite{nakamoto2008peer}, 
has a variant~\cite{sasson2014zerocash} that can ensure the validity and privacy of transactions without a centralized agency that is trusted by everyone, by utilizing two cryptographic tools called commitment and zero-knowledge proof~\cite{de1987non}. Concepts of these cryptographic tools have also been adopted to design new proposals to address global governance issues, such as nuclear arms-control~\cite{glaser2014zero}.

In this article, we convert the data dilemma of balancing data authenticity and data security in a multi-country ETS to a sequence of information hiding and binding problems. We demonstrate that a cryptographic system can address these problems and build mathematically verifiable trust without compromising the security of sensitive data. 

\section{From system requirements to cryptographic problems}

\subsection{Overview of the emissions accounting system}
We describe a stylized emissions accounting system to facilitate our discussion on addressing the data dilemma. This setting might not perfectly match current emissions accounting practices adopted in different jurisdictions, but we expect key features we consider here are in line with the trend of technologies and rules that are ideal for the future multi-country ETS.

We focus on interactions between firms in a specific country of the multi-country ETS (hereafter firms), the government of the country (hereafter the country), an authorized international auditing committee (hereafter the verifier). We first assume that firms are all large stationary sources, ensuring that randomly verifying a meaningful subset of firms by on-site checks is feasible. We further assume that there are continuous emissions monitoring meters deployed at all the stacks of covered firms that release CO$_2$ into the atmosphere. There are internationally-agreed standards that these meters should comply with to collect granular emissions data (e.g., CO$_2$ concentration and flow rate in every hour). These granular data can be used to calculate firm-level aggregated emissions within a compliance cycle. Once collected, the granular emissions data should be stored in a secure local repository that could not be tampered with and stand ready for possible verification. Firms then privately report their aggregated emissions data to the country and are allowed not to disclose the aggregated data. The country then sums up firm-level emissions within their boundaries and publishes the aggregated emissions at the country-level. The country-level emissions are important for tracking the progress towards achieving the country’s individual nationally determined contributions (NDCs) under the Paris Agreement.

A random verification mechanism is embedded in the system. We assume that the country provides a full list of firms within their boundaries and agrees that a random subset of firms can receive on-site checks with their granular and firm-level emissions data exposed to the verifier for close examination. Firms thus face a valid threat of being caught if they manipulate the meters to fudge granular emissions data or misreport any data.

The following features need to be guaranteed by cryptography to ensure data authenticity and security of this accounting system.
\begin{itemize}
\item 	The randomness of the firm list for the verification is accepted by all the parties.
\item	For firms selected for the verification, their reported data are verifiable, meaning that data misreporting in any step of the reporting process can be detected.
\item 	For firms not selected for the verification, their granular and aggregated emissions data are invisible to others. 
\item 	It is verifiable that the country has honestly reported the country-level emissions. 
\end{itemize}

\subsection{The cryptographic protocol design problem}
We translate the above requirements for the system into a cryptographic protocol design problem described as follows. 

\begin{enumerate}
\item Firm $i$ collects granular emissions data in short time intervals (e.g., hours), measured by meters that meet an internationally agreed standard. Its emissions in a certain hour $h$ is $e_{i,h}$.  All the granular emissions data are attached with a digital signature from the meter so that the firm can not tamper with the original data. The emissions data and signature are then stored in a secure local repository that can prevent tampering or breaking. The firm calculates its total emissions within a compliance cycle (e.g., a year), $\ane_i$, by summing up all the hourly emissions ($\ane_i=\sum_h e_{i,h}$). The firm then converts $\ane_i$ into a committed message $c_i$ and publishes it. The firm privately reports $\ane_i$ as well as necessary inputs in computing $c_i$ to the country.
\item The country conducts a check to ensure that all the $c_i$ are computed correctly and then publishes the aggregated country-level emissions $\ane$, where $\ane=\sum_i\ane_i$. 
\item The country also provides a full list of indexed firms to the verifier. The verifier and the country collaboratively generate a random subset of firms for the verification. If firm $i$ is selected, it is subject to examination on the meter operation and reporting process to check whether $e_{i,h}$ is correctly generated and signed. The firm should also disclose the $\ane_i$ and necessary inputs in computing $c_i$. The verifier will check if $c_i$ is generated correctly. 
\end{enumerate}
	
It becomes clear that we need to design the protocols for the following three cryptographic tasks, all of which are cryptographic primitives that have been well studied~\cite{goldreich2009foundations}.

\begin{itemize}
\item \textbf{(Hiding and binding)} When a firm converts its emission data $\aneb_i$ to committed message $c_i$, others cannot infer $\aneb_i$ from $c_i$ (hiding). Once submitted, the firm cannot specify another different value $\aneb’ _i\neq \aneb_i$ that can be converted to the same $c_i$ (binding)\footnote{We consider a situation that the firm follows the algorithm in converting $\aneb_i$ to $c_i$ but may report $\aneb_i$ not the same as the real data $\ane_i$.}.

\item \textbf{(Random list generation)} The verifier and the country jointly choose a subset of firms. Neither party has the capability to influence the probability of being selected for any firm.

\item \textbf{(Verifiable summation) } The verifier can check whether $\ane$ is computed correctly, even not all the $\ane_i$ are visible to the verifier. 
\end{itemize}

\section{Constructing protocols}

In this section, we use $[k]$ to denote the set $\{x\in \mathbb{Z}|1\le x\le k\}$, where $k$ is an integer. 
Cryptography relies on a special mathematical tool called the \emph{elliptic curve group} to construct variety protocols. 
An elliptic curve group consists of a set of points on an elliptic curve (e.g., $y^2=x^3+x+7$).\footnote{An elliptic curve group also contains a special point called the infinity point. Here we omit it for simplicity.} 
An additive operation is defined on these points.\footnote{The additive operation has the communicative property and associative property.} 
For a given point $P$ on an elliptic curve, we use $P+P$ to denote the result of the additive operation and use $a\cdot P$ for $a\in \mathbb{Z}^+$ to denote $\underbrace{P+\cdots+P}_{\text{repeat }a\text{ times}}$.
There exists a special starting point $G$, from which a sequence of points $G,2\cdot G,3\cdot G,\cdots$ can be generated with the following properties:
\begin{enumerate}
    \item $n\cdot G$ is periodic in $n$ and its period is a large prime integer $q$, which means $n\cdot G=(n+q)\cdot G$.
    \item For a random $n$, it is hard to infer $n$ from $n\cdot G$, meaning that the inference will take decades using modern computer resources. 
\end{enumerate}

The cryptography literature \cite{lochter2010elliptic,qu1999sec,langley2016rfc} has identified concrete choices for elliptic curves and starting point $G$ that can generate an elliptic curve group.

\subsection{Commitment}
With the elliptic curve group, a \emph{commitment scheme} can be constructed~\cite{pedersen1991non}. Intuitively, a commitment scheme resembles a security safe. A committer can put some data in this safe and transfer it to the others. Unless the committer provides the key for the safe, the others can not learn the data inside. This property can be referred as the \emph{hiding property}. After the committer transfers the safe to the others, the committer can no longer change the data inside. This property is referred as the \emph{binding property}.

Formally, with chosen elliptic curve and the starting point $G$, the commitment scheme consists of three algorithms:
\begin{itemize}
    \item \emph{Setup}: Someone samples a secret random value $h$ from $[q-1]$ and computes the public value $H=h\cdot G$, meaning that $H$ should be revealed to the committer but $h$ should be kept secret from the committer. (Recall Property 1 of the elliptical curve group that $q$ is the period of the sequence $G,2\cdot G,\cdots$.)
    \item \emph{Make commitment} (put data in the safe): In order to commit an integer $0\le m < q$, the committer chooses a random value $r$ from $[q]$ and computes the commitment value $c=m\cdot G+r\cdot H$.\footnote{For any point $P$, $0\cdot G + P=P$. Essentially, $0\cdot G$ is the infinity point.} We use $f_H(m,r)$ to denote $m\cdot G + r\cdot H$ hereafter. 
    \item \emph{Reveal commitment} (open the safe): When a committer is required to open the safe, the committer should reveal $m$ and $r$ and the equality $c=f_H(m,r)$ will be checked. 
\end{itemize}

On the one hand, the above algorithms can achieve the \emph{hiding property}. Note that $f_H(m,r)=m\cdot G+r\cdot H=(m+r\cdot h)\cdot G$. Since $r$ is random, $m + r\cdot h \mod q$ is also random according to the elementary number theory. The randomness of $f_H(m,r)$ hides information about $m$. 

On the other hand, the algorithms can also achieve the \emph{binding property}. The committer cannot tamper the data after commitment. Property 2 of the elliptic curve group states that the committer cannot infer $h$ from $H$, so it further guarantees that the committer cannot find $m'\neq m$ and another $r'$ that satisfy $f_H(m,r)=f_H(m',r')$ to tamper the data when revealing the commitment.\footnote{If the committer finds such $m'$ and $r'$, we must have $f_H(m,r)=f_H(m',r')$ and further $(m+rh)\cdot G=(m'+r'h)\cdot G$. Since the sequence of $n\cdot G$ is periodic with a period of $q$, $q$ is divisible by $(m'+r'h)-(m+rh)$. The committer can then learn $h$ by computing $h=(r'-r)^{-1}\cdot (m'-m) \mod q$.}

\subsection{Hiding and binding}

The \emph{commitment scheme} is utilized as the cornerstone of our proposed emissions accounting system. Each firm $i$ uses the function $f_H$ to convert the firm-level emissions $\ane_i$ into a secure safe $c_i$ that contains the information of $\ane_i$ and publish it. It can be formalized by the following \emph{hiding and binding protocol}. 

$\\$\noindent\textbf{Hiding and binding protocol}
\begin{enumerate}
    \item The verifier makes setup for the commitment and publishes $H$. 
    
    \item The firm $i$ makes the commitment by picking random $r_i$ and computes $c_i=f_H(\ane_i,r_i)$. 
\end{enumerate}

    This \emph{hiding and binding protocol} guarantees that others cannot learn emissions data of firm $i$ after reporting, unless the firm is selected for the verification. During the verification process, the verifier will learn $\ane_i$ and $r_i$ as well as $e_{i,h}$. She will verify whether $e_{i,h}$ has been correctly generated by examining the status of the meter, whether $\ane_i$ is correctly calculated from $e_{i,h}$, and whether $c_i$ is correctly calculated from $\ane_i$ and $r_i$. Firm $i$ cannot disclose a different aggregated emissions value from $\ane_i$ after submission because the firm is not possible to find another pair of $\ane_i$ and $r_i$ that produces $c_i$.

\subsection{Random list generation}

It is a prerequisite for our emissions accounting system that the verifier and the country should reach a consensus on a randomly selected subset of firms for the verification. Neither the verifier nor the country should have the capacity to influence the random list generation. Here we design a \emph{random value generation protocol} based on the commitment scheme for the \emph{random list generation protocol} described as below.

$\\$\noindent\textbf{Random list generation protocol}
\begin{enumerate}
    \item The country provides a full list of firms to be selected. 
    \item Let $L$ be the length of the firm list, the country and the verifier generate a random value $m$ from $[L]$ using a \emph{random value generation protocol}.
    \item  Add the $m$-th firm to the verification list and eliminate it from the firm list.
    \item The above two steps are repeated $k$ times to add $k$ firms in the verification list.
\end{enumerate}

The \emph{random value generation protocol} asks the verifier and the country to provide a random value $m_v$ and $m_c$, respectively. If no cryptographic algorithm is applied and $m_v+m_c\mod L$ is selected naively as the random value $m$, the randomness cannot be guaranteed. This is because the country can create a preferred value $\tilde{m}$ by choosing $\tilde{m}_c=\tilde{m}-m_v$ if the country learns $m_v$ before it chooses $m_c$ or is able to tamper with $m_c$ after it is chosen. Hence, we also need to adopt the \emph{commitment scheme} to design the \emph{random value generation protocol}. 

We let the country and the verifier put their choices ($m_v$ and $m_c$) in security safes ($c_v$ and $c_c$) first, then swap the safes and open the safes. After swapping the safes and before opening them, the country and the verifier have locked their choices in the safes and have not seen the other's choice, so each party's choice must be independent from the other's. This process is formalized as following steps:

$\\$\noindent\textbf{Random value generation protocol}
\begin{enumerate}
    \item The verifier and the country run the setup process for commitment and publish $H_v$ and $H_c$, respectively.  
    \item The verifier chooses a random value $m_v$, and makes commitment by picking $r_v$ and computing $c_v=f_{H_c}(m_v,r_v)$. The verifier then publishes $c_v$. The country computes $c_c$ and publishes it in the same way.
    \item The verifier publishes $m_v$ and $r_v$, so the country can check whether $c_v$ is correctly calculated using $m_v$ and $r_v$. The country does the same thing. 
    \item Let $m=m_v+m_c\mod L$ be the output random value.
\end{enumerate}

The \emph{random value generation protocol} guarantees that as long as one party follows the protocol and submits random value ($m_v$ or $m_c$), the output $m$ must be randomly distributed. Hence, the verification list generated following the \emph{random list generation protocol} is guaranteed to be random.

\subsection{Verifiable summation}

Our emissions accounting system needs to ensure that the country correctly computes the country-level emissions $\ane$ from firm-level emissions data $\{\ane_i\}_{i=1}^n$, which have been converted to committed messages $\{c_i\}_{i=1}^n$. Since the \emph{commitment scheme} naturally holds the additive property, \footnote{$f_H(m_1,r_1)+f_H(m_2,r_2)=(m_1+m_2)\cdot G + (r_1+r_2)\cdot H=f_H(m_1+m_2,r_1+r_2)$} $\sum_i c_i$ is essentially the commitment of data $\sum_i E_i$ with the random value $\sum_i r_i$. Formally, we design a \emph{verifiable summation protocol} to achieve the mechanism.

$\\$\noindent\textbf{Verifiable summation protocol}
\begin{enumerate}
    \item The country receives $\{\ane_i\}_{i=1}^n$ and $\{r_i\}_{i=1}^n$ from all the firms and publishes $\ane=\sum_i \ane_i$ and $r=\sum_i r_i$. 
    \item The verifier checks whether $\sum_i c_i=f_H(E,r)$. 
\end{enumerate}

\subsection{Review}

\begin{figure}[bth]
\raggedleft
        \includegraphics[width=1.3\linewidth,center]{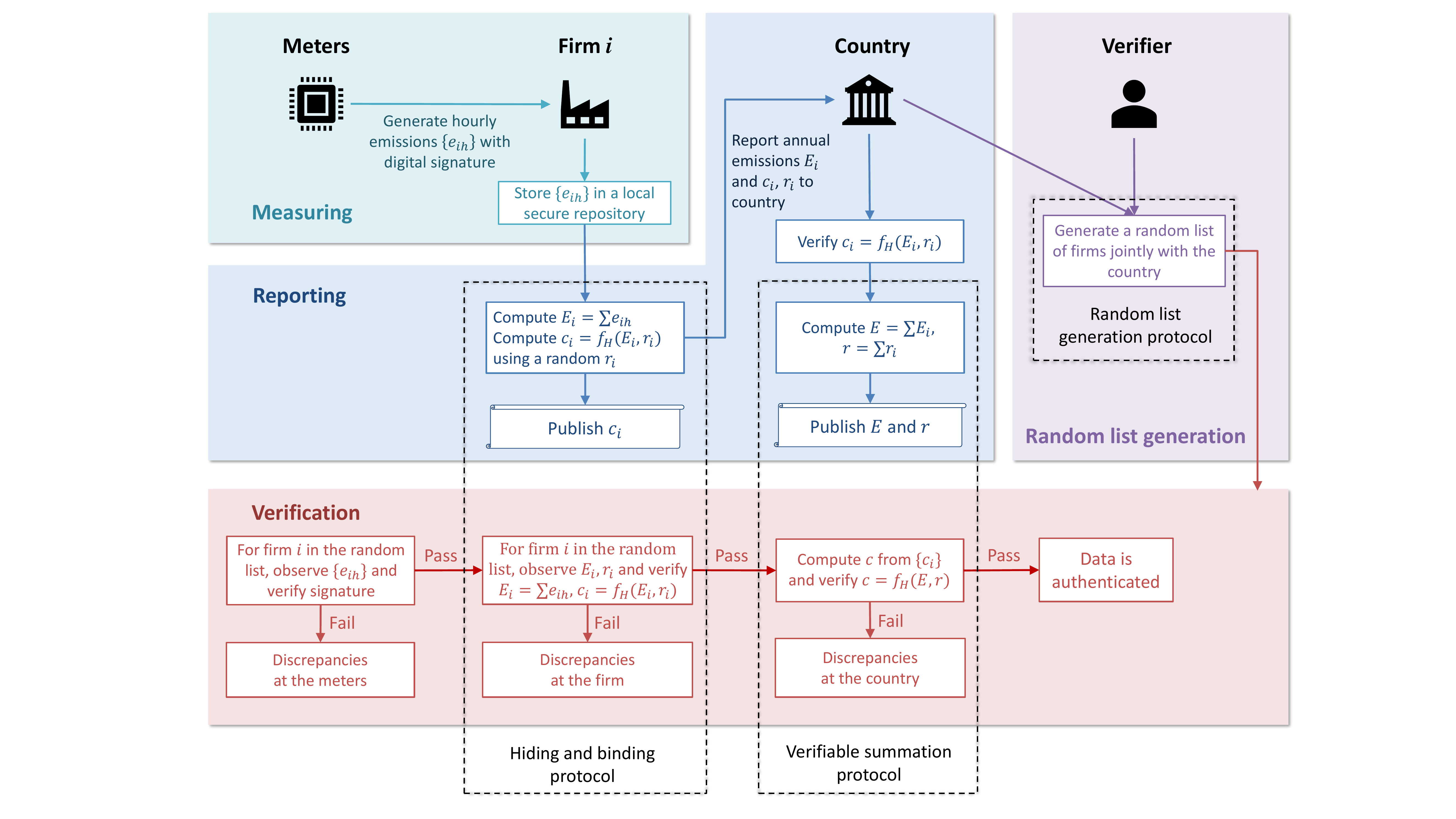}
        \caption{Overview of the emissions accounting framework.}
        \label{fig:framework}
    \vspace{0em}
\end{figure}

Figure~\ref{fig:framework} summarizes the emissions accounting framework by linking all the protocols to the measuring, reporting, and verification process.

\section{Discussion}
The authenticated and secure emissions accounting system that we have presented so far provides a cryptographic framework that can preserve data authenticity and security for a stylized multi-country ETS. Several additional aspects are worth discussion.

First, our framework highlights the role of international standards for the emissions metering system, including sensors, auxiliary hardware, and digital signature schemes. These standards ensure the authenticity of granular emissions data, on which the whole reporting and verification processes are built. Reducing the cost of the system is crucial to enable its mass deployment because high-accuracy CO$_2$ sensors are so far still costly.

Second, although we confine our framework within interactions between parties from a single country in the multi-country ETS and the verifier for the clarity of presentation, existing cryptographic protocols can extend the \emph{random list generation protocol} to multiple countries~\cite{schindler2020hydrand}. All the reporting and verification processes discussed above are still compatible with reasonable extensions when parties from different countries are interacting at the same time with the verifier. 

Third, some more complicated cryptographic tools might be added to the framework in the real-world application. For instance, we might need to introduce a specific form of \emph{zero-knowledge proof protocol}, called \emph{range proof protocol}\cite{lipmaa2003diophantine,bunz2018bulletproofs}, to avoid some undesirable behaviors. In an extreme scenario, the country and one of the firms may collude to substantially reduce the country-level emissions $E$ by letting the firm report a negative firm-level emissions $E_i$. Unless this specific firm is selected in the random verification, the manipulation will not be detected by the \emph{verifiable summation protocol} as the summation of country-level emissions is done honestly by the country. The \emph{range proof protocol} can be applied to only allow a plain message $E_i$ that is an integer between 0 and a reasonably large value for firm-level emissions to be submitted.

The \emph{zero-knowledge proof protocol} can be further incorporated to strengthen our system. In our current design, the verifier still needs to check the sensitive data for selected firms, and firms still have chance to misreport without being caught. These two limitations can be addressed by adopting a \emph{zero-knowledge proof protocol} that consists of a prove algorithm and a verification algorithm in the \emph{hiding and biding protocol}. The prove algorithm takes all the sensitive data $e_{i,h}, E_i$, the message that contains information about a firm's self-reported emissions, $c_i$, and $r_i$ as inputs and outputs a proof that contains no information about the sensitive data. The verification algorithm takes the committed message and the proof as inputs, and can judge whether the firm's self-reported message is generated faithfully.

We acknowledge that many more data aspects that are crucial for constructing future multi-country ETS have been left out of our discussion. However, we believe the framework we sketch in this article has demonstrated some desirable features that modern cryptography can bring along to the future international climate governance, especially facilitating the negotiation and implementation of Article 6 of the Paris Agreement. We urge more research in this direction to strengthen the trust and accelerate collaborative actions to achieve climate targets.

{\footnotesize \bibliographystyle{acm}
	\bibliography{paper}}
\clearpage

\appendix
\newcommand{\stp}{\mathsf{Setup}}
\newcommand{\comm}{\mathsf{Commit}}
\newcommand{\view}{\mathsf{View}}
\newcommand{\audit}{\mathsf{audit}}
\newcommand{\pick}{\mathsf{pick}}
\newcommand{\stat}{\mathsf{stat}}
\newtheorem{define}{Definition}

\newenvironment{proof}{\vspace{-0.05in}\noindent{\bf Proof.}}%
{\hspace*{\fill}$\Box$\par\vspace{0.4cm}}

\end{document}